\documentclass[intlimits,twoside,a4paper]{article}

\usepackage{amsmath,amssymb}
\usepackage{graphicx}

\usepackage[T2A]{fontenc}
\usepackage[cp1251]{inputenc}

\usepackage{cmpj2}

\issue{2012}{15}{2}{23301}

\doinumber{10.5488/CMP.15.23301}

\newcommand{\HB}{\rm HB}

\title[Percolation and Widom lines]%
{Percolation line, response functions, and Voronoi polyhedra analysis in supercritical water\thanks{Dedicated to Professor Orest Pizio on the occasion of his 60th birthday.}}
\author[J. \v{S}kvor, I. Nezbeda]
{J. \v{S}kvor\refaddr{label1}, I. Nezbeda\refaddr{label1,label2}
}
\addresses{
\addr{label1} Faculty of Science, J.E. Purkinje University,
 400 96 \'{U}st\'{\i} nad Labem, Czech Republic
\addr{label2} Institute of Chemical Process Fundamentals, Academy of Sciences,
165 02 Prague 6, Czech Republic
}

\authorcopyright{J. \v{S}kvor, I. Nezbeda, 2012}
\date{Received February 19, 2012, in final form March 21, 2012}

\begin{document}

\maketitle

\begin{abstract}

The problem of a physical relevance (meaning) of percolation in
supercritical fluids is addressed considering a primitive model of water. Two different criteria, physical and configurational,
are used for the cluster definition in Monte Carlo simulations over a
range of pressures to determine the percolation line and skewness, and
a theoretical analytic equation of state is used to evaluate response
functions. It is found that both criteria yield practically the same
percolation line. However, unlike the findings for simple fluids, the
loci of the response function extrema exhibit density/pressure
dependence quite different from that of the percolation line. The only
potential coincidence between the loci of the extrema of a
thermodynamic property and a detectable structural change is found for
the coefficient of isothermal compressibility and Voronoi neighbors distribution skewness maximum.
\pacs 36.40
\end{abstract}


\section{Introduction}

It has been well known that at certain thermodynamic conditions some
properties of fluids may exhibit a rather sudden sharp change. Typical
examples are a rapid change of electric conductivity of aqueous
solutions~\cite{BonnEPhysL58,LalEPhysJE15}, metal-nonmetal transition
in supercritical fluids~\cite{CouletJChP118}, or a sudden change of
electric and dielectric properties of composites~\cite{NanReview}.
Formation of clusters (aggregates) of molecules and/or a network has
been made responsible for these observations.

Unlike the crystalline phase, studies of percolation in a fluid phase
encounter a number of both technical and fundamental (methodological)
problems. First of all, in continuous systems with a continuous
interaction potential it is not clear how to define a bond. Hence, several
criteria have been put forward~\cite{SatorPR376,SengerJChP110}.
Similarly, using
molecular simulations on finite samples the question arises how to
define and detect the presence of an infinite cluster. Consequently,
practically no theoretical results are available for fluids while various
conjectures on percolation in fluids represent a mere extension of the
results  originally obtained for lattice systems. We have addressed
these problems in a series of recent papers~\cite{SkvorPRL,SkvorCCCC,SkvorPRE} with the following two most important
results. It was shown that (i) unlike the lattice systems, the
percolation in continuous systems does not follow any scaling law and
cannot be thus characterized by universal constants and that (ii) the
wrapping probability (the cluster wraps the system if, starting from any particle of the cluster and moving along
interparticle bonds, there is a possibility of
getting to an image of that particle in another replica~\cite{SeatonJChP86}) provides a more accurate estimate of the
percolation threshold compared to the crossing probability (the
cluster crosses the system if the maximal distance between some pairs
of its particles is equal to or greater than the box length)~\cite{SkvorPRL,SkvorCCCC}.

Instances of the observed sharp changes in some properties of fluids
usually refer to mixtures. However, it is also well known that the
response functions (derivatives of thermodynamic potentials) of
pure fluids exhibit extrema in the supercritical region~\cite{XuPNAS102}. It means that
some thermodynamic properties change the course of their dependence on
the thermodynamic conditions and the natural question to be asked is
what is behind this change in the behavior at the molecular level. A
potential answer is that these changes may be associated with the
formation of an infinite cluster. If this is the case, then the
percolation line, i.e., a line connecting all state points where the percolation transition/threshold occurs, becomes a quantity with a real physical content and
will not represent a mere mathematical construction extending otherwise
meaningful results.

In a recent paper~\cite{SkvorMP} we have addressed this important issue
by studying (by Monte Carlo simulations) the percolation and various
structural characteristics of two simple fluids, the square-well and
Lennard-Jones ones. We have found that out of all the considered extrema lines of response functions  (RFEL; i.e., the lines connecting the state points at which the given response function, as a function of temperature $T$ at constant pressure $P$, reaches its extremum) only the RFEL of the coefficient of isothermal compressibility has  closely followed the course of the skewness extremum line (SEL; i.e., the line connecting the state points at which the skewness of distribution of the number of nearest neighbors in terms of the Voronoi tessellation~\cite{MedvedevJCP67,MontoroJChP97,JedlovszkyJChP111} reaches its maximum) whereas no definite conclusion has been possible to draw concerning other response functions.
A similar study has been conducted recently by Idrissi et al.~\cite{IdrissiJPChB115} who investigated the relation of thermodynamic response functions with the Voronoi tessellation properties in supercritical ammonia. They found a potential relation between the  thermal expansion coefficient and other Voronoi tessellation properties, e.g., the volume distributions. Specifically, the volume distributions broaden near the temperature where the thermal expansion coefficient reaches its maximum. Nonetheless, the relation between the percolation transition on one hand and the thermodynamic response functions and/or Voronoi tessellation on the other hand has not been fully explored yet.

Accounting for a number of uncertainties when dealing with simple fluids the inconclusive results mentioned above are not surprising. As a more severe test of the original surmise of a coincidence of the percolation and SEL lines and the RFEL lines, a system with a natural formation of chains or clusters should be considered. Water is a typical example of a system with association and we have therefore carried out the same project considering a simple model of water, i.e., the primitive model, for which (i) the criterion of bonding is unique and (ii) for which an analytic equation of state is also available~\cite{Vlcek_EOS}.

In the present paper we report the results for the same properties as in the previous study~\cite{SkvorMP}. In the next section we provide all necessary definitions and computational details and in section 3 we present and discuss the obtained results.

\section{Basic definitions and computational details}

In this paper we consider a primitive model of water (PMW) descending from the realistic parent TIP4P model hereafter referred to as the PM/TIP4P model~\cite{VNPM2}.
This is a simple short-range archetype of real water  faithfully reproducing its structure and results from a non-speculative modeling applied to the parent model~\cite{PCCPrev}. In the approach used, the Lennard-Jones
interaction in the realistic parent model is replaced by a hard-sphere
repulsion, while the repulsive and attractive electrostatic interactions
at short separations are approximated by hard sphere repulsion and
square-well attraction, respectively (for details see~\cite{VNPM2,VNPM0}). Thus, the functional form of the model reads as
\begin{eqnarray}
\label{uPM}
u_{\rm PM}(1,2)&=&u_{\rm HS}(r_{\rm OO},d_{\rm OO})
  + \mathop{\sum_{i\in\{1\},j\in\{2\}}} u_{\rm HS}(r_{ij};d_{ij})
  + \mathop{\sum_{i\in\{1\},j\in\{2\}}} u_{\rm SW}(r_{ij})\nonumber\\
&=&u_{\rm PHB}(1,2)
  + \mathop{\sum_{i\in\{1\},j\in\{2\}}}u_{\rm SW}(r_{ij}),
\end{eqnarray}
where
\begin{equation}
\begin{array}{rcll} u_{\rm HS}(r;d) & = & +\infty, &
\text{for} \qquad r<d,
\\ & = & 0, & \text{for} \qquad  r>d,
\end{array}
\end{equation}
\begin{equation}
\begin{array}{rcll}
u_{\rm SW}(r;\lambda ) & = & -\epsilon_{\rm HB}, & \text{for} \qquad
r<\lambda,
\\ & = & 0, & \text{for} \qquad r>\lambda,
\end{array}
\end{equation}
and the summation in the second term of (\ref{uPM}) runs over the pairs
of the like sites and in the third term of the unlike sites. The above PMW was
developed to serve as a short-range reference system for developing a
full perturbation theory of water~\cite{JirsakN_EOS}. It was shown to
faithfully reproduce the structure of water as well as to capture a good
deal of its thermodynamic properties including the observed anomalies.
The parameters of the model are presented in table~\ref{tab1}. It is worth recalling
that the potential $u_{\rm PHB}$ in (1) incorporates all repulsive
interactions of the PMW; after switching off all attractive
interactions in the PMW we are left with a rather strange hard body
called pseudo-hard body (PHB)~\cite{PHB1}. It is not a simple hard body
because it possesses a flavor of nonadditivity and currently no theory
for the fluid of PHB's is available\cite{PHB2}.

\begin{table}[h]
\caption{Parameters of the considered PM/TIP4P model of water~\cite{VNPM2}.\label{tab1}}
\vspace{2ex}
\centering
\begin{tabular}{l l}
\hline
parameter&$[d_{\textrm{OO}}]$\\
\hline
$d_{\textrm{XX}}$ & $0.8$\\
$\lambda_{\textrm{XM}}$ & $0.65$\\
$r_{\textrm{OX}}$ & $0.5$\\
$r_{\textrm{OM}}$ & $0.06$\\
\hline
\end{tabular}
\end{table}

For the PMW with its strongly orientation dependent short-range
attractive interaction, it is only natural to consider two molecules to be
bonded when their interaction energy $u_{ij}$ is $-\epsilon_{\HB}$ (the
parameters of the model satisfy the so-called conditions of saturation
which means that no more than one bond can be established between a
pair of molecules). The other criterion used to define a bond between
two molecules is the energy-based Hill's definition according to which
two molecules are bonded when their relative kinetic energy is less
than their interaction energy~\cite{SatorPR376,HillJChP23}.
Specifically, if ${\bf v}_i$ and ${\bf v}_j$ are the velocity vectors of
particles $i$ and $j$, respectively, and $m$ is their mass, then they are
considered to be bonded if
 \begin{equation}
 \frac{m}{4}\left({\bf v}_j - {\bf v}_i\right)^2 \leqslant - u_{ij}\,.\label{HillCriterion}
\end{equation}

We used the common Metropolis Monte Carlo (MC) simulation in an NVT ensemble~\cite{AllenBook} both for the location of the percolation line and SEL which means that two types of simulations had to be performed. The wrapping probability~$R_\mathrm{w}$ as a function of reduced number density~$\rho^*$ at a given temperature~$T$ and number of particles~$N$ have to be determined first. Then, the percolation threshold density~$\rho^*_\mathrm{c}(T)$ is calculated as the density where $R_\mathrm{w}$ at given~$T$ for different $N$ intersect in one point (more accurate numerical procedure is described in~\cite{SkvorPRE}). Details of this simulation were as follows: (i) $N$ was set to $1000$, $2000$, or $4000$, (ii) each $R_\mathrm{w}(\rho^*,\{N,T\}={\rm const})$ was determined for at least 22 (with maximum up to 42) densities $\rho^*$ with step $0.002$ such that the minimum of the mean of that $R_\mathrm{w}$ was less than $0.03$ and its maximum was greater than $0.97$, (iii) the systems were equilibrated by performing at least $32\times10^4 N$ MC steps, (iv) equilibrium configurations used for the detection of the presence of the wrapping cluster were separated by $3N$ MC steps, (v)~at least $7\times10^6$ configurations (up to $124\times10^6$; increasing with $T$ going closer to critical point and with smaller $N$) were investigated to reach a sufficiently low error of measurement, and (vi) to apply the energy criterion, the velocity components were randomly assigned to every particle from the Gaussian distribution characterized by the temperature of the system (for further details see the previous paper~\cite{SkvorPRE}). The simulation details for determining the SEL were as follows: (i) $N$ was set to $1000$, (ii) each skewness $\gamma_1(\rho^*,T={\rm const})$ was determined for at least 28 values of $\rho^*$ (with step $0.002$), (iii) the systems were equilibrated by performing at least $10^5 N$ MC steps, and (iv) $10^6$ equilibrium configurations, separated by $100N$ MC steps, were used for the Voronoi polyhedra analysis in which the water molecules were represented only by the position of their oxygen-like sites. The parameters of all the simulations mentioned above were set so as to maintain the acceptance ratio around~1/3.

The response functions that we consider to be of interest are the coefficient of thermal expansion,
\begin{equation}
\alpha_P = \frac{1}{V}\left(\frac{\partial V}{\partial T}\right)_P
 = -\frac{1}{\rho}\left(\frac{\partial\rho}{\partial T}\right)_P  \,\,,
\label{alphaP}
\end{equation}
the coefficient of isothermal compressibility,
\begin{equation}
\kappa_T =-\frac{1}{V}\left(\frac{\partial V}{\partial P}\right)_T =
\frac{1}{\rho}\left(\frac{\partial\rho}{\partial P}\right)_T  \,\,,
\end{equation}
and the isobaric heat capacity,
\begin{equation}
c_P = \frac{1}{N}\left(\frac{\partial H}{\partial T}\right)_P  \,\,,
\label{cP}
\end{equation}
where $H$ is enthalpy. Finally, an extension of the vapor-liquid (V-L) coexistence curve to the supercritical
temperatures was determined by extrapolating the Clausius-Clapeyron equation $\ln P=a-b/T$, where $a$ and $b$ are fitted parameters for the model considered. Both the response functions (and corresponding RFEL's) and the extension of V-L line were determined numerically from Vl\v{c}ek and Nezbeda equation of state~\cite{Vlcek_EOS}.
An analytic theory of the considered PMW was developed using a variant of the 2nd
order thermodynamic perturbation theory (TPT2)~\cite{TPT2} which was
shown to be very accurate. In general, the residual Helmholtz energy
per particle, $a^{\rm res} = A^{\rm res}/N$ for the fluids consisting
of molecules with $m$ hydrogen-like sites and $n$ complementary $N$-sites
assumes the form~\cite{Vlcek_EOS,JirsakN_EOS}
  \begin{equation}
  \beta a_{\rm PM}^{\rm res} = \beta a^{\rm res}_{\rm PHB} + m(1-\nu) + \ln x_0   \,\,,
   \label{APM}
  \end{equation}
where $x_0$ is given, using the reduced density, $\rho^*=\rho d_{\rm
OO}^3$, by
  \begin{equation}
  x_0=\frac{\nu^m}{\left[1 + m{\rho^*}\nu I_{+-}
   + (m{\rho^*}\nu)^2I_{+-+}\right]^n}\,\,,
  \label{x_0}
  \end{equation}
and parameter $\nu$, measuring `non-saturation', satisfies the
following cubic equation:
  \begin{eqnarray}
  m^2{\rho^*}^2I_{+-+}\nu^3 & + & \left[\left(2mn-m^2\right){\rho^*}^2I_{+-+}
  + m{\rho^*} I_{+-}\right]\nu^2 \nonumber\\
  &+&  \left[(n-m){\rho^*} I_{+-} + 1\right]\nu - 1 = 0.
  \label{e1}
  \end{eqnarray}
Quantities $I$ are integrals involving the correlation function of the
underlying PHB fluid and are available in a parametrized  form (for
further details and values of the parameters see reference~\cite{JirsakN_EOS}).
For compressibility factor, $z=\beta P/\rho$, one gets the following
result
  \begin{eqnarray}
  z& = & \rho\left(\frac{\partial \beta a^{\rm res}_{\rm PM}}{\partial \rho}\right)_\beta +1 \nonumber \\
  &=& z_{\rm PHB} + \rho\left(\frac{\partial}{\partial \rho} \left[m(1-\nu) + \ln x_0\right] \right)_\beta\, .
  \label{EOSreal}
  \end{eqnarray}

\section{Results and discussion}

\begin{figure}[!t]
\centerline{\includegraphics[width=0.75\textwidth]{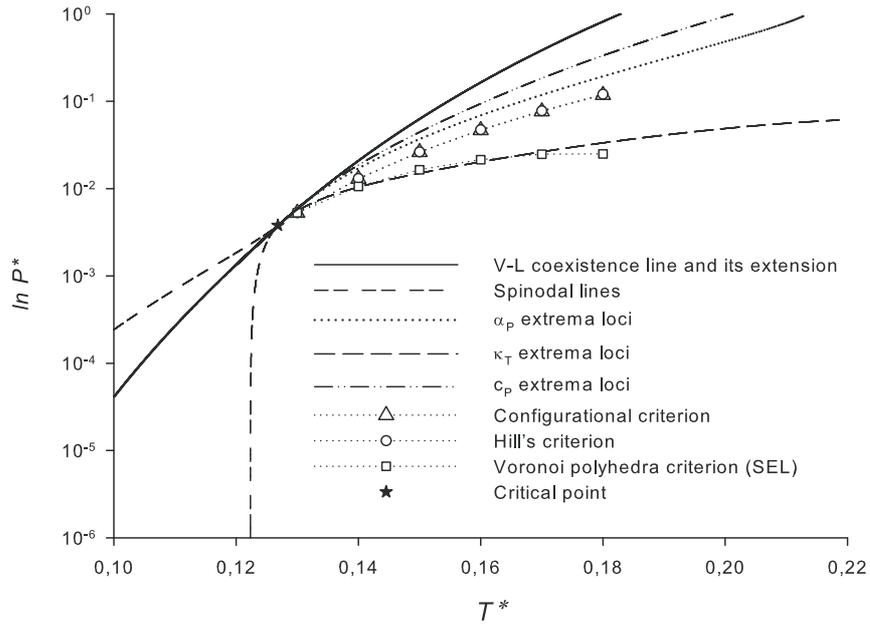}}
\caption{The percolation lines, Voronoi tessellation skewness extremum line, and response function extrema lines in the supercritical PM/TIP4P fluid in the pressure-temperature plane.}\label{fig1}
\end{figure}

\begin{figure}[!b]
\centerline{\includegraphics[width=0.8\textwidth]{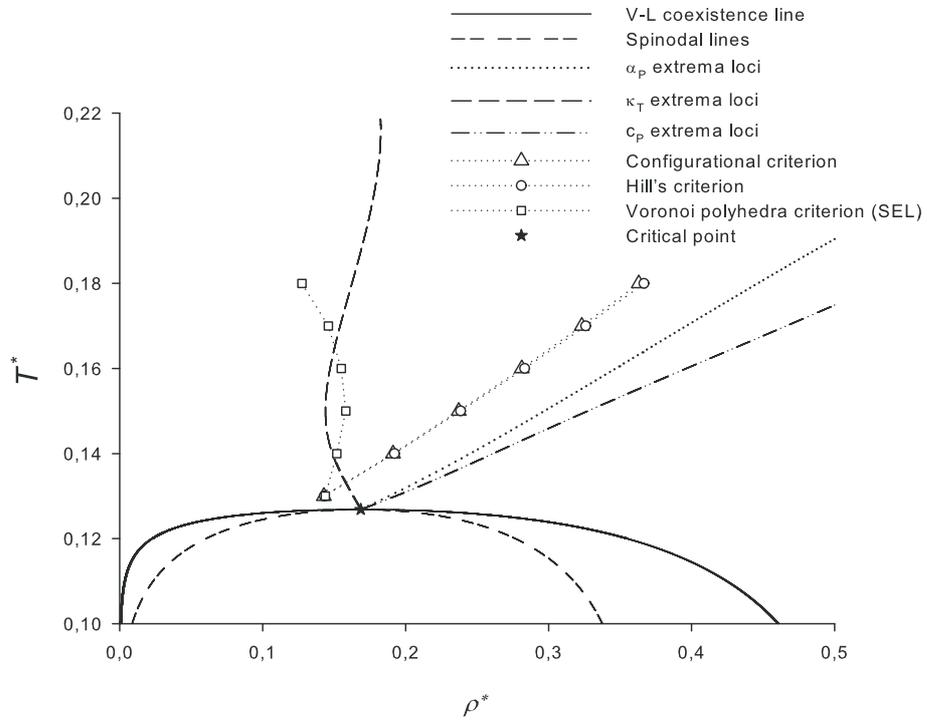}}
\caption{The same as figure~\ref{fig1} in the temperature-density plane.}\label{fig2}
\end{figure}

Following the strategy outlined above using simulations we determined two types of the percolation line corresponding to the two different definitions of bonds (configurational and physical) and the results are shown in figures~\ref{fig1} and~\ref{fig2}.
The first finding is discernible. The configurational and physical criteria yield the identical percolation line. In our previous papers we always used these two criteria because they always produced different lines giving rise to an ambiguity in the interpretation of the results. The reason why this is not the case for the considered model probably lies in the choice of thermodynamic conditions: unlike in the previous studies, we operate in the low temperature supercritical region. Stillinger's criterion imposes that $u_{ij}<0$. Hill's criterion is stronger, it requires that $u_{ij}+k_{ij}<0$ where $k_{ij}>0$ is the relative kinetic energy. Since the two percolation lines are practically identical, it means that the relative kinetic energy of two molecules is negligible in comparison with the H-bonding energy. This argument is supported by the results obtained previously for the same model at higher temperatures. Due to a high temperature, the kinetic energy attains large values and starts competing with the H-bond energy leading to different conditions for percolation and thus to two different percolation lines.

Unlike in our previous papers on percolation, where we investigated this phenomenon over a wide range of thermodynamic conditions, in this paper we have focussed on supercritical temperatures as close to the critical point as possible. The still open question whether the percolation line originates/termi\-nates in the critical point is one of other important issues. Examination of the figures leads to the result similar to that obtained for simple fluids: in the $P-T$ plane, the percolation line tends  to the critical point as predicted by theory but this is not the case in the $T-\rho$ plane. However, when discussing this issue one has to bear in mind that while the model and the range of the considered temperatures have removed the ambiguity concerning the definition of bonding of two molecules, a certain uncertainty yet remains. Namely, evaluation of the model properties from an equation of state. The used equation was shown to be quite accurate at high temperatures but its accuracy deteriorates with decreasing temperature. Particularly, the location of the critical point determined from this equation is only approximate.

As usual, we show the percolation lines and the RFEL' in two different frames: in the $P-T$ plane which is common when discussing any transition and in the $T-\rho$ plane which is a common practice when dealing with percolation. As we see, results in both cases are a bit different and we attribute this difference to the use of only an approximate equation of state as mentioned above.
To discuss the results for the response functions, it is useful to distinguish two types of the response functions: (i) one, reflecting the geometrical arrangement of the molecules, i.e., the coefficient of isothermal compressibility, and (ii) the remaining functions that involve energy.
 In full agreement with the previously obtained results, we find that
the coefficient of isothermal compressibility follows the course of the SEL. This is correct if we examine this behavior in the $P-T$ plane, whereas in the $T-\rho$ plane there is only a rough similarity in the course of these two functions. As regards other RFEL's, the result is similar to that found for the simple fluids: the percolation line follows the course which differs from that of the RFEL, although qualitatively there may be found the same functional dependence.

The final remark on the obtained results concerns an 'imaginary' extension of the vapor-liquid (V-L) coexistence curve also shown in the figures. The relation between the percolation line and the extended coexistence curve in supercritical water was discussed already by P\'{a}rtay and co-workers~\cite{PartayJChP123,PartayJPChB111}. In the first paper they pointed out a possible coincidence of both lines and in the second one they modified their claim stating only that the percolation line goes through higher pressures than the extension of the V-L line. However, quite an opposite conclusion can be drawn from the present paper. Taking into account that we used a different model (primitive contrary to realistic one) and that the extended coexistence curve has no physical sense, such disagreement is not that surprising. Moreover, the extension of the V-L line was already shown~\cite{SkvorMP} to have no relation either to any of the RFEL's or to the percolation line in the case of simple fluids and the same holds true for the considered model of water.

\section{Conclusions}

Water molecules naturally associate and the definition of the bond is unique for the model considered. We have, therefore, embarked on this project with the hope to find at least some definite answers to the questions raised in the Introduction. It means, to find a link between the percolation and/or certain structural characteristics of the fluid and the observed changes in the thermodynamic behavior of supercritical fluids. The only positive, and probably generally valid result is that the rearrangement of molecules under pressure and witnessed by the  Voronoi neighbors distribution skewness maximum gives rise to an extremum of the coefficient of isothermal compressibility. Concerning other response functions, we have to admit that the achieved results turned out to be disappointing leaving the question of the physical sense of the percolation in supercritical fluids unanswered, the same applying to the origin of the observed extrema of the response functions.

\section*{Acknowledgement}

This work was supported by the Grant Agency of the Academy of Sciences of the Czech Republic (Grant No.~IAA200760905) and of the Grant Agency of the J.E.~Purkinje University (Project No.~53223--15--0010--01). Valuable comments of an anonymous referee are also greatly acknowledged.

\newpage
\ukrainianpart

\title{Лінія перколяції, функції відгуку і аналіз багатогранника Вороного в надкритичній воді}
\author[]
{Я. Шквор\refaddr{label1}, І. Незбеда\refaddr{label1,label2}
}
\addresses{
\addr{label1} Факультет природничих наук, Університет Я.Е. Пуркінйє, Усті над Лабем, Чеська Республіка
\addr{label2} Інститут фундаментальних основ хімічних процесів, Академія наук, Прага, Чеська Республіка
}

\makeukrtitle

\begin{abstract}
\tolerance=3000%
Проблема фізичної важливості (суті) перколяції у надкритичних плинах
досліджується на прикладі примітивної моделі води.  Два різних
критерії, фізичний і конфігураційний, використовуються для означення
кластера в моделюванні методом Монте Карло при різних тисках, з метою
отримання перколяційної лінії та її перекосу. Теоретичне аналітичне
рівняння стану використовується для оцінки функцій відгуку.  Знайдено,
що обидва критерії дають практично однакову лінію перколяції. Проте,
на відміну від результатів для простих плинів, геометричне місце точок
екстремумів функції відгуку вказує на цілком іншу залежність від
тиску/густини, ніж лінія перколяції.  Єдине потенційне співпадіння між
геометричним місцем точок екстремумів термодинамічної характеристики і
спостережуваної структурної зміни є знайдено для коефіцієнта
ізотермічної стисливості і максимуму асиметрії розподілу Вороного.
\keywords лінія перколяції, функції відгуку, лінії Відома, надкритична вода, мозаїка Вороного
\end{abstract}

\end{document}